\documentstyle[prl,aps
]{revtex}
\begin{document}

\twocolumn[\hsize\textwidth\columnwidth\hsize\csname@twocolumnfalse%
\endcsname

\preprint{SU-ITP \#94/38, cond-mat/9411029}

\title{ {\hfill\normalsize SU-ITP \#94/38, cond-mat/9411029\medskip\\}
        Universal Fluctuation of the Hall Conductance \\
        in the Random Magnetic Field}

\author{Kar\'en Chaltikian, Leonid Pryadko and Shou-Cheng Zhang}
\address{ Department of Physics, Stanford University, Stanford, CA
  94305}

\date{January 25, 1995}

\maketitle

\begin{abstract}
  We show that the RMS fluctuation of the antisymmetric part of the
  Hall conductance of a planar mesoscopic metal in a random magnetic
  field with zero average is universal, of the order of $e^2/h$,
  independent of the amplitude of the random magnetic field and the
  diffusion coefficient even in the weak field limit. This quantity is
  exactly zero in the case of ordinary scalar disorder. We propose an
  experiment to measure this surprising effect, and also discuss its
  implications on the localization physics of this system.   Our
    result applies to some other systems with broken time-reversal
    ({\bf T}) symmetry.
\end{abstract}

]

Recently, the problem of electron's motion in a random magnetic field
has been investigated both theoretically and experimentally. The main
motivation stems from the mapping of strongly correlated electron
systems to problems of fermions coupled with gauge fluctuations. In
particular, it has been argued that the DC transport properties of the
half filled Landau level can be mapped through a Chern-Simons
transformation to a problem of fermions moving in a static random
magnetic field\cite{kz,hlr,kwaz}.  The configuration of static random
magnetic field can also be realized directly in
experiments\cite{geim,nordita,mancoff}.  In the recent experiment by
Mancoff {\em et al.}~\cite{mancoff} a giant, $V$-shaped
magneto-resistance was found in the system of two dimensional electron
gas placed on top of a demagnetized permanent magnet.  This experiment
verified the connection between the DC transport at half filled Landau
level and the problem of electron's motion in a random magnetic field.
The large magnitude of the magneto-resistance could raise some
possibilities in technological applications.

{}From theoretical point of view, the effect of quantum interference
on the transport behavior in these systems is extremely interesting.
It is well known that quantum interference generally gives rise to
localization in two dimensions. An important exception is the two
dimensional electron gas in a uniform quantizing magnetic field, where
extended states exist at the center of each Landau band.  Hall
conductance plays an important role in the localization properties of
a two dimensional electron gas. In the language of the effective field
theories, the Hall conductance leads to an extra topological term in
the non-linear sigma model with unitary symmetry~\cite{pruisken} and
it has been argued that this term is responsible for the existence of
extended states. In view of the effect of the Hall conductance on the
localization properties, it is natural to ask about the Hall
conductance in the case of the random magnetic field. For random
magnetic fields with zero average, the average of the Hall conductance is
certainly zero by the symmetry, but the variance of the Hall
conductance may still have some non-trivial properties.

Within the classical theory of transport, the local value of the Hall
conductance is proportional to the local magnetic field and the
dissipative conductance, so that its RMS fluctuation
depends on material characteristics and scales inversely with the
square root of the system size. Hall conductance is determined by the
amplitude of the local magnetic field and for weak fields never
reaches the quantum values.  This argument appears to imply that
fluctuations of the Hall conductance in the random flux problem in the
ohmic regime are unimportant and this system is very different from
the case of a uniform magnetic field where the Hall conductance is of
the order of $e^2/h$.

However, it is known that quantum interference plays a significant
role in mesoscopic systems and leads to universal fluctuations of
transport coefficients\cite{leestone,altshuler}. At low temperatures
when the inelastic mean free path exceeds the system size, the system
does not self-average. At zero temperature, the fluctuations of
transport coefficients persist even in the infinite system limit, and
more over, these fluctuations are universal and independent of the
material characteristics.

In this paper, we compute explicitly the RMS fluctuation of the
antisymmetric part of the Hall conductance
$\tilde\sigma_{xy}=(\sigma_{xy}-\sigma_{yx})/2$ in the random flux
problem.  We show that because of the quantum interference effects,
this quantity is universal,
\begin{equation}
  \langle\!\langle\tilde\sigma_{xy}^2\rangle\!\rangle \sim
  \left({e^2}/{h}\right)^2.
\label{eq3}
\end{equation}
Such a behavior is remarkably different from what one would expect
classically. As long as the system remains mesoscopic,~(\ref{eq3})
holds independent of the system size, although it depends weakly on
geometry. The large value of RMS implies that the typical values of
$\tilde\sigma_{xy}$ are comparable to the Hall conductance in
quantizing uniform magnetic field. This result is highly surprising
since it holds in the limit where the random magnetic field is weak.
We shall discuss its interpretation and implications at
the end of the paper.

Before we present the details of our calculations, we would like to
remark on an important technical issue. The fluctuation of the
transverse conductance $\sigma_{xy}$ in mesoscopic samples has been
computed before for the orthogonal ensemble. Ma and Lee~\cite{ma-lee}
found that $\sigma_{xy}$ has universal fluctuation in the absence of
any external magnetic field.  This fluctuation comes from the fact
that there is no $90^0$ rotation symmetry in a given member of the
orthogonal ensemble, so that a current in the $X$ direction can be
scattered randomly into the $Y$ direction.  Fluctuations of
$\sigma_{xy}$ have the same origin as fluctuations of $\sigma_{xx}$,
and are not related to the {\bf T}-symmetry breaking. A better
definition of the Hall conductance in mesoscopic samples is the
antisymmetric part $\tilde\sigma_{xy}$.  It can be shown that
$\tilde\sigma_{xy}$ vanishes identically for every member of the
orthogonal ensemble.   The external magnetic field $B_0$ breaks
  down {\bf T}-symmetry, resulting in non-zero fluctuations of
  $\sigma_{xy}^a(B_0)=(\sigma_{xy}(B_0)-
        \sigma_{xy}(-B_0))/2$~\cite{LKD,Webb}.
  In the systems with purely potential disorder
  $\tilde\sigma_{xy}=\sigma_{xy}^a$ due to the Onsager relation.  When
   disorder breaks the {\bf T}-symmetry, $\tilde\sigma_{xy}$ and
  $\sigma_{xy}^a$ become very different. In particular, $\sigma_{xy}^a$
  vanishes identically in zero external field while
  $\tilde\sigma_{xy}$ does not.

Following Ma and Lee~\cite{ma-lee}, we start from Str{\v
  e}da~\cite{streda} formula for the Hall conductivity
$\sigma_{xy}=\sigma_{xy}^{\rm I}+\sigma_{xy}^{\rm II},$
\begin{eqnarray} \label{sigmaone}
  \sigma_{xy}^{\rm I}&=&\frac{i\hbar e^2}{2} {\rm Tr} \left[v_x
  G^{+}_{E_F} v_y \rho(E_F) - v_x \rho(E_F) v_y G^{-}_{E_F} \right],\\
\label{sigmatwo}
\sigma_{xy}^{\rm II}&=&-\sigma_{yx}^{\rm II}= ec\frac{\partial
  N(E_F)}{\partial B},
\end{eqnarray}
where $G^{\pm}(E)=\left(E-\hat{H}\pm i\eta\right)^{-1}$ is the
retarded (advanced) Green's function, $N(E)=S^{-1}{\rm
  Tr}\int_{-\infty}^E\rho(\epsilon)d\epsilon$ is the specific number
  of states and \begin{equation} \label{rho-of-E}
  \rho(E)=i/(2\pi S)\left(G^+_E-G^-_E\right)
\end{equation}
is the operator of the density of states; we assume the wave functions
normalized to one on the whole area $S$ of the sample.  The second
part of the Hall conductivity is already antisymmetric,
$\sigma_{xy}^{\rm II}\equiv \tilde\sigma_{xy}^{\rm II};$ it is
identically zero in the case of potential scattering in zero external
magnetic field and  its average is small in weak enough external
  magnetic field as long as $\langle N(E_F)\rangle$ has almost no
  dependence on $B_0$. Nevertheless, interference effects result in
  quite significant value of the RMS fluctuation of $\sigma_{xy}^{\rm
    II}$ despite the wide-spread belief of its smallness.

There are several important length scales in the problem. We denote
$l_c$ the correlation length of the magnetic field, $B_{\rm rms}$ the
RMS value of the random magnetic field and $r_c=v_F/\omega_c=m v_F
c/(e B_{\rm rms})$ the classical cyclotron radius. The effect of the
magnetic field is strong if it dramatically changes the trajectories
of the particles; this is certainly the case if $r_c\lesssim l_c$ and
electron may form a closed orbit inside a single correlated domain of
magnetic field.

All our calculations will be carried out in the opposite, weak field
regime where $r_c\gg l_c$, so that within a domain the trajectories of
the particles are only slightly bent by the magnetic field.  If the
magnetic field fluctuations are still slow enough compared to
electron's de Broglie wavelength, $\lambdabar= k_F^{-1} \ll l_c,$ the
scattering of the electron wave packet by the magnetic field is a
quasiclassical process.  Neglecting the small correlations arising
when the particle returns to the same point via different trajectories
(these are important for the physics of localization,) one can easily
understand such a classical motion: it is a diffusion with coefficient
\begin{equation}\label{diffusion-constant}
  D=\frac{1}{2}\frac{\tau_{\rm tr} v_F^2}{1+\Omega_0^2\tau_{tr}^2},
\end{equation}
where $\tau_{tr}\sim r_c^2/(v_F l_c)$ and $\Omega_0=eB_0/mc$ is the
cyclotron frequency associated with the uniform part of the external
magnetic field $B_0.$ It is amazing that
equation~(\ref{diffusion-constant}) holds also in the opposite limit
$l_c\ll \lambdabar$ of quantum diffraction of electrons on the
magnetic field fluctuations; the only difference is that now
$\tau_{tr}={2\hbar}/{m l_c^2\langle\omega_c^2\rangle }$~\cite{mirlin}.
The localization corrections arise at long distances due to the
multiple scattering {}from the same regions of the magnetic field; in
mesoscopic systems their effects are small as long as
$\lambdabar/l_{tr}\sim\hbar/(E_F\tau_{tr})\ll 1.$ This relationship is
generally valid assuming $r_c\gg l_c;$ we believe that this last
condition determines the validity of our results.

In performing perturbation calculations for the random flux problem,
one has to be very careful in always computing gauge invariant
quantities.  The perturbation expansion is ill-defined for
gauge-dependent quantities like the averaged one-particle Green's
function; they are non-analytic as a function of small expansion
parameter $\lambdabar/ l_{tr}$. The gauge-invariant quantities like
the density of states $\rho(E)$ or diffusion propagator are, however,
free of divergences and can be evaluated
systematically~\cite{mirlin,mirlin-two,km}.

In the usual case of weak scalar disorder there are two major
long-range effective modes: Diffuson and Cooperon. Only Diffuson
survives in the magnetic disorder problem because its existence is
granted by the conservation of the number of particles; all other
normal modes decay at the distances of the order of the mean free path
$l_{tr}$. As usual, Diffuson corresponds to an average of the product
of advanced and retarded one-particle Green's functions
\begin{equation}\label{diffusion}
  \Gamma(q,w)\equiv \overline{G^+({\bf p}\!+\!{\bf
      q},E\!+\!iw)G^-({\bf p},E)}\propto {1\over{i\omega-Dq^2}}.
\end{equation}
The diffusion propagator $\Gamma({\bf q},\omega)$ depends on the
uniform part $B_0$ of the magnetic field mainly through the diffusion
coefficient (\ref{diffusion-constant}). However, if the retarded and
advanced propagators in~(\ref{diffusion}) are evaluated at the
different values of the magnetic field, say $B_1$ and $B_2$, the
amplitudes for direct and return paths no longer cancel and the
diffusion propagator acquires the form~\cite{lee-stone-f}
\begin{equation}\label{diffusion-gauge}
  \Gamma(q,w)\propto \left({i\omega-D\left(\hat{\bf q}+{e}{\bf
      a}/\hbar c\right)^2}\right)^{-1},
\end{equation}
where the vector potential ${\bf a}$ corresponds to the difference
between the two magnetic fields, $\nabla \times {\bf {a}}=B_1-B_2.$

Starting with the magnetization part~(\ref{sigmatwo}) of the Hall
conductivity, we rewrite the density of particles
\begin{equation}
  N(E)= \frac{i}{2\pi S}\!\int_{-\infty}^E\!\! d\epsilon {\rm
    Tr}\left(G_{\epsilon}^+-G_{\epsilon}^-\right),
\end{equation}
and integrate over $\epsilon$ explicitly to obtain
\begin{displaymath}
  \left\langle\!\!\left\langle\!  {\left.\!\sigma_{xy}^{\rm
        II}\!\right.}^2 \right\rangle\!\!\right\rangle\! =- e^2c^2
    \frac{\partial^2}{\partial B_{1,2}} \frac{
      \left\langle\!\left\langle (\ln\! G^+_1\!-\!\ln\!  G^-_1) (\ln
        \!G^+_2\!-\!\ln\!  G^-_2) \right\rangle\!\right\rangle
        }{4\pi^2S^2},
\end{displaymath}
where $G^{\pm}_1$ and $G^{\pm}_2$ are the non-averaged Greens
functions in the presence of magnetic fields $B_1$ and $B_2$
respectively. These fields are introduced here to define the Hall
conductance; after taking the derivatives in the above equation their
difference should be set to zero.  In the diagrammatic language, the
irreducible average in the above equation can be represented as a sum
of averaged products of electron's vacuum loops.  Neglecting weak
localization effects, each term can be evaluated as a vacuum diagram
made out of the single two-particle correlator
\begin{equation}
  \unitlength=1mm \left\langle\!\!\!\left\langle
\begin{picture}(16.00,8)(2.0,9.0)\scriptsize\thicklines
  \put(10.00,10.00){\circle{14.00}} \put(10.00,10.00){\circle{6.00}}
  \put(06.20,16.0){\makebox(0,0)[cc]{x}}
  \put(13.80,16.00){\makebox(0,0)[cc]{x}}
  \put(17.00,10.00){\makebox(0,0)[cc]{x}}
  \put(03.00,10.00){\makebox(0,0)[cc]{x}}
  \put(06.20,4.00){\makebox(0,0)[cc]{x}}
  \put(13.80,4.00){\makebox(0,0)[cc]{x}}
  \put(08.1,12.20){\makebox(0,0)[cc]{x}}
  \put(11.9,12.20){\makebox(0,0)[cc]{x}}
  \put(12.90,10.00){\makebox(0,0)[cc]{x}}
  \put(07.30,10.00){\makebox(0,0)[cc]{x}}
  \put(08.20,7.80){\makebox(0,0)[cc]{x}}
  \put(11.80,7.80){\makebox(0,0)[cc]{x}}
  \put(10.00,4.50){\makebox(0,0)[cc]{\large $\scriptstyle +$}}
  \put(10.00,9.00){\makebox(0,0)[cc]{\large $\scriptstyle -$}}
  \put(10.00,17.1){\makebox(0,0)[cc]{$>$}}
  \put(10.00,12.76){\makebox(0,0)[cc]{$<$}}
\end{picture}
\right\rangle\!\!\!\right\rangle\! =
\begin{picture}(16.00,9)(2.0,9.0)\scriptsize \thicklines
  \put(10.00,10.00){\circle{14.00}} \put(10.00,10.00){\circle{6.00}}
  \put(06.20,16.0){\makebox(0,0)[cc]{x}}
  \put(13.80,16.00){\makebox(0,0)[cc]{x}}
  \put(17.00,10.00){\makebox(0,0)[cc]{x}}
  \put(03.00,10.00){\makebox(0,0)[cc]{x}}
  \put(06.20,4.00){\makebox(0,0)[cc]{x}}
  \put(13.80,4.00){\makebox(0,0)[cc]{x}}
  \put(08.1,12.20){\makebox(0,0)[cc]{x}}
  \put(11.9,12.20){\makebox(0,0)[cc]{x}}
  \put(12.90,10.00){\makebox(0,0)[cc]{x}}
  \put(07.30,10.00){\makebox(0,0)[cc]{x}}
  \put(08.20,7.80){\makebox(0,0)[cc]{x}}
  \put(11.80,7.80){\makebox(0,0)[cc]{x}}
  \put(10.00,4.50){\makebox(0,0)[cc]{\large $\scriptstyle +$}}
  \put(10.00,9.00){\makebox(0,0)[cc]{\large $\scriptstyle -$}}
  \put(10.00,17.1){\makebox(0,0)[cc]{$>$}}
  \put(10.00,12.76){\makebox(0,0)[cc]{$<$}}
  \small \multiput(8.1,11.8)(-0.25,0.5){8}{\makebox(0,0)[cc]{$\cdot$}}
  \multiput(11.9,11.8)(0.25,0.5){8}{\makebox(0,0)[cc]{$\cdot$}}
  \multiput(7,10)(-0.6,0.0){8}{\makebox(0,0)[cc]{$\cdot$}}
  \multiput(13,10)(0.6,0.0){8}{\makebox(0,0)[cc]{$\cdot$}}
  \multiput(11.9,8.2)(0.25,-0.5){8}{\makebox(0,0)[cc]{$\cdot$}}
  \multiput(8.1,8.2)(-0.25,-0.5){8}{\makebox(0,0)[cc]{$\cdot$}}
\end{picture}.
\label{trivloop}
\end{equation}
Since only the simultaneous product of logarithms of retarded and
advanced electron's Green functions produces a long-range Diffuson, we
get
\begin{equation}
  \langle\!\langle (\sigma_{xy}^{\rm II})^2\rangle\!\rangle=-
  \left.\frac{e^2c^2}{2\pi^2S^2} \frac{\partial^2}{\partial
      B_{1,2}} {\rm Tr} \ln D\left(\hat{\bf q}+{e{\bf a}\over\hbar
      c}\right)^2\right|_{b=0}.
\label{unitary-case}
\end{equation}
Neglecting the weak dependence of the diffusion
coefficient~(\ref{diffusion-constant}) on the magnetic field, and
noticing that now everything depends only on the difference
${b}=B_1-B_2,$ we obtain the general answer
\begin{equation} \label{final-gtwo}
  \langle\!\langle (\sigma_{xy}^{\rm II})^2\rangle\!\rangle=+
  \left.\frac{e^2 c^2}{2\pi^2S^2} \frac{\partial^2}{\partial b^2}
    {\rm Tr} \ln \left(\hat{\bf q}+{e{\bf a}\over\hbar
      c}\right)^2\right|_{b=0}.
\end{equation}
We evaluate this sum inside a box with zero boundary conditions in
$y$-direction and periodic boundary conditions in $x$-direction.
After expansion in powers of ${\bf a}$ to the second order, the
r.~h.~s. of~(\ref{final-gtwo}) becomes
\begin{equation}\label{boxed}
  \frac{e^4}{2\pi^2\hbar^2}\frac{1}{S^2}\left[
  \sum_{\alpha}{\frac{\langle\alpha|y^2|\alpha\rangle}{q_\alpha^2}}
  -2\sum_{\alpha\beta} {\frac{\left|\langle\alpha|y
      q_x|\beta\rangle\right|^2}{q_{\alpha}^2 q_{\beta}^2}} \right].
\end{equation}
Here $\alpha$ and $\beta$ label the normalized wave functions and
$q_{\alpha,\beta}^2$ are the corresponding eigenvalues of the square
of the wave vector.  Each of two terms in~(\ref{boxed}) diverge at
infinity, but their difference converge. The result may be written as
$\langle\!\langle (\sigma_{xy}^{\rm
  II})^2\rangle\!\rangle=\left({e^2}/{h}\right)^2 g(\beta), $ where
$g(\beta)$ is a universal function depending on the dimensionless
geometrical parameter $\beta=L_x/L_y.$ For the square sample $L_x=L_y$
we obtain $\langle\!\langle (\sigma_{xy}^{\rm II})^2\rangle\!\rangle=
0.060 ({e^2}/{h})^2.$

Again, we emphasize the difference between the pre\-sent system and
the orthogonal ensemble, where one has an additional contribution from
the Cooperon channel and the equation (\ref{unitary-case}) has the
form
\begin{displaymath}
  \langle\!\langle (\sigma_{xy}^{\rm II})^2\rangle\!\rangle
  \propto \frac{\partial^2}{\partial B_{1,2}} {\rm Tr} \left[ \ln
  \left(\hat{\bf q}+{e{\bf a}\over\hbar c}\right)^2 + \ln
    \left(\hat{\bf q}+{e{\bf A}\over\hbar c}\right)^2\right]
\end{displaymath}
with $\nabla \times {\bf A}=B_1+B_2$. It is easy to see that
$\langle\!\langle (\sigma_{xy}^{\rm II})^2\rangle\!\rangle$
vanishes identically in the orthogonal case.

In the situation of mixed disorder the Cooperon may be not suppressed
at higher momenta, setting an upper cut-off on the summation
in~(\ref{boxed}). However, since the sum is convergent, our result is
still valid when the magnetic disorder dominates.

Now consider the conventional part~(\ref{sigmaone}) of the Hall
conductivity. With~(\ref{rho-of-E}), it's antisymmetric part becomes
\begin{equation}\label{g-one}
  \tilde\sigma_{xy}^{\rm I}=\frac{\hbar e^2}{4\pi}{\rm Tr}\left(\hat
  v_xG_{E_F}^+\hat v_yG_{E_F}^--\hat v_xG_{E_F}^-\hat
  v_yG_{E_F}^+\right).
\end{equation}
The irreducible average of the square of this quantity should contain
only diffusion propagators as shown
\begin{displaymath}
  \setlength{\unitlength}{0.75mm}
  \left\langle\!\!\!\left\langle\left(
\begin{picture}(20,10)(10,9)\scriptsize
  \thicklines \put(10,10){\circle*{1.5}}
  \put(30,10){\circle*{1.5}}
  \put(20,10){\oval(20,10){}}
  \put(20,15){\makebox(0,0)[cc]{$>$}}
  \put(20,05){\makebox(0,0)[cc]{$<$}}
  \multiput(14,15)(04,00){4}{\makebox(0,0)[cc]{x}}
  \multiput(14,05)(04,00){4}{\makebox(0,0)[cc]{x}}
  \put(12.5,12){\makebox(0,0)[cc]{$+$}}
  \put(12.5,08){\makebox(0,0)[cc]{$-$}}
\end{picture}
\,-\,
\begin{picture}(20,10)(10,9)\scriptsize
  \thicklines \put(10,10){\circle*{1.5}}
  \put(30,10){\circle*{1.5}}
  \put(20,10){\oval(20,10){}}
  \put(20,15){\makebox(0,0)[cc]{$>$}}
  \put(20,05){\makebox(0,0)[cc]{$<$}}
  \multiput(14,15)(04,00){4}{\makebox(0,0)[cc]{x}}
  \multiput(14,05)(04,00){4}{\makebox(0,0)[cc]{x}}
  \put(12.5,12){\makebox(0,0)[cc]{$-$}}
  \put(12.5,08){\makebox(0,0)[cc]{$+$}}
\end{picture}
\right)^{\!\!2}\,\right\rangle\!\!\!\right\rangle= 2 \times
\begin{picture}(30,10)(5,9)\scriptsize
  \thicklines \put(10,10){\circle*{1.5}} \put(30,10){\circle*{1.5}}
  \put(07,10){\circle*{1.5}} \put(33,10){\circle*{1.5}}
  \put(20,10){\oval(26,16){}} \put(20,10){\oval(20,08){}}
  \put(20,14){\makebox(0,0)[cc]{$<$}}
  \put(20,06){\makebox(0,0)[cc]{$>$}}
  \put(20,18){\makebox(0,0)[cc]{$>$}}
  \put(20,02){\makebox(0,0)[cc]{$<$}}
  \multiput(14,14)(04,00){4}{\makebox(0,0)[cc]{x}}
  \multiput(14,06)(04,00){4}{\makebox(0,0)[cc]{x}}
  \multiput(14,18)(04,00){4}{\makebox(0,0)[cc]{x}}
  \multiput(14,02)(04,00){4}{\makebox(0,0)[cc]{x}}
  \multiput(14,14)(04,00){4}{
    \multiput(0,0)(0,0.8){5}{\makebox(0,0)[cc]{\small $\cdot$}}}
  \multiput(14,02)(04,00){4}{
    \multiput(0,0)(0,0.8){5}{\makebox(0,0)[cc]{\small $\cdot$}}}
  \put(12.5,12){\makebox(0,0)[cc]{$+$}}
  \put(12.5,08){\makebox(0,0)[cc]{$-$}}
  \put(07,16){\makebox(0,0)[cc]{$-$}}
  \put(07,04){\makebox(0,0)[cc]{$+$}}
\end{picture}.
\end{displaymath}
For the usual disorder there is also a negative contribution due
to Cooperons, suppressing Hall conductance to zero.
Formally, the diagram above is equivalent to the diagram (3b) in
Ref.~\cite{ma-lee} without the factor $4$ since our definition of
$\tilde\sigma_{xy}$ includes $1/2.$ We obtain~\cite{ma-lee-note}
\begin{equation}
  \langle\!\langle (\tilde\sigma_{xy}^{\rm
    I})^2\rangle\!\rangle\approx 0.411/4\, (e^2\!/h)^2 = 0.103
    (e^2\!/h)^2.
\end{equation}
 It is easy to see that the cross term
  $\langle\!\langle\tilde\sigma_{xy}^{\rm I} \tilde\sigma_{xy}^{\rm
    II}\rangle\!\rangle$ is zero for weak enough external magnetic
  fields.
Combining the results for $\tilde\sigma_{xy}^{\rm I}$ and
$\tilde\sigma_{xy}^{\rm II},$ we finally obtain
\begin{equation}
  \langle\!\langle \tilde\sigma_{xy}^2\rangle\!\rangle\approx0.16\,
  (e^2\!/h)^2.
\end{equation}

This result can also be derived more systematically from the
non-linear $\sigma$-model formulation of the random flux problem; due
to limitation in space we will not present the details here.

Even though all our perturbative calculations are carried out in the
weak field regime $l_c \ll r_c$, the RMS fluctuation in Hall
conductance is of the order of $e^2/h$, a value which is typically
seen in the strong field regime. In this sense, our result is highly
surprising. Our physical picture of this result is the following: for
a given member of the random flux ensemble, each energy eigenstate can
be assigned with an integer Chern number, defined as the Hall
conductance in units of $e^2/h$ averaged over the twisted boundary
conditions~\cite{kohmoto}.  In the orthogonal case, the time-reversal
symmetry ensures that all eigenstates are real and have zero Chern
numbers.  For the random flux problem, if the Fermi energy lies below
the mobility edge, all states are localized and have zero Chern
numbers.  However, for Fermi energy above the mobility edge, which
always exists in metallic samples of finite size, states extended
beyond the edges of the sample have finite Chern numbers. While the
probability for either sign is the same for a given state, these Chern
numbers are correlated over an energy range of $\Delta E= \hbar D/S$.
At high temperature, if the inelastic broadening of the energy levels
is much greater than $\Delta E$, one measures small Hall conductance
due to the randomness of signs. However, for mesoscopic systems, the
level broadening is typically of the order of $\Delta E$, the
correlated structure of the Chern numbers becomes measurable. As one
varies the Fermi energy, the RMS fluctuation of the Chern numbers is
of the order of one.

Direct experimental verification of this result has to be carried out
in the regime where
$$ l_{\phi}=\sqrt{D\tau_{\phi}}\gg L_{x,y} \gg l_{tr}.
$$ The first condition ensures that one has a mesoscopic system while
the second one results in suppression of the Cooperons (here $l_{tr}$
is the mean free path determined by the scattering on the magnetic
field fluctuations). The magnetic field fluctuations do not have to be
very short-ranged as long as condition $r_c\gg l_c$ is also satisfied.
One has to perform a four probe measurement in order to be able to
extract the antisymmetric part of the conductivity, for example, by
interchanging the current and the voltage leads. The necessary
sampling can be achieved by varying the Fermi energy, the gate
voltage,  or the external magnetic field. The correlation of
the values measured at different Fermi energies or different magnetic
fields decay in the same fashion as the usual mesoscopic fluctuations.

It is of course highly interesting to explore the consequences of this
universal fluctuation on the localization properties in this system.
If an average Hall conductance of the order of $e^2/h$ is important
for the existence of extended states in two dimensions, what about
this universal fluctuation of the Hall conductance of the same order?
In fact, Zhang and Arovas~\cite{za} have argued recently that the
long-ranged fluctuations of the Hall conductance in this system give
rise to a long ranged interaction between the topological densities,
and this new extra term could lead to extended states. The existence
of the extended states in the random flux problem is still an on-going
debate, and different views are held by various
groups~\cite{kz,kwaz,za,debate-pro,debate-con}.  Our result reveals
the microscopic origin of the long ranged topological interaction
discussed in reference \cite{za}. Since the typical value of the Hall
conductance is of the order of $e^2/h$, it is plausible that extended
state might exist in the random flux problem for reasons similar to
those in the case of quantizing uniform magnetic field.

 Let us comment on the applicability of our results to other
  systems with broken {\bf T}-invariance.  We used only two properties
  of the electron's motion in the random magnetic field: the diffusion
  propagator of the form~(\ref{diffusion}) and the suppression of the
  Cooperon.  The same features characterize electron's motion in the
  systems with usual spin disorder or even with usual scalar disorder
  in the presence of weak ($\hbar/mS\lesssim\Omega_0\ll 1/\tau_{tr}$)
  uniform magnetic field.  The first inequality provides for the
  suppression of the Cooperon while the second one ensures that the
  magnetic field is not quantizing.

We would like to acknowledge interesting and fruitful discussions with
D.~Arovas, M.~Dykman, M.~Ma, C.~Marcus and P.~A.~Lee.  Part of this
work was carried out while S.~C.~Z. was vising the Hong Kong
University of Science and Technology.\ \ L.~P. is thankful to IBM for
support through the IBM Graduate Fellowship.

After completing this letter, we received a preprint~\cite{newest}
where the RMS fluctuation of $\sigma_{xx}$ in the random flux problem
was found to be universal.


\begin{references}
\bibitem{ma-lee-note} In the reference \cite{ma-lee} the result for
  $g_2\approx 0.37$ is inaccurate by ten percent. We use our value
  $g_2\approx 0.411$.
\end{references}
\end{document}